\documentstyle[prb,aps,epsf,floats]{revtex}

\begin{document}

\draft

\twocolumn[\hsize\textwidth\columnwidth\hsize\csname
@twocolumnfalse\endcsname

\title{Hall-effect in LuNi$_2$B$_2$C and YNi$_2$B$_2$C 
borocarbides: \newline a comparative study}

\author{V. N. Narozhnyi$^{a,b,c}$\cite{PresAdd}, J. 
Freudenberger$^b$, V.N. Kochetkov$^{a,c}, $ K. A. 
Nenkov$^{b,c}$\cite{OnLeave}, G. Fuchs$^b$, A.~Handstein$^b$,
K.-H.~M\"uller$^b$}

\address{$^a$Institute for High Pressure Physics, Russian Academy of 
Sciences, Troitsk, Moscow Region, 142092, Russia} 
\address{$^b$Institut f\"ur Festk\"orper- und Werkstofforschung 
Dresden e.V., Postfach 270016, D-01171 Dresden, Germany}
\address{$^c$International Lab. of High Magnetic Fields and Low
Temperatures, Gajowicka 95, 53-529 Wroclaw, Poland}

\date{\today}

\maketitle

\begin{abstract}
The Hall effect in LuNi$_2$B$_2$C and YNi$_2$B$_2$C borocarbides has 
been investigated in normal and superconducting mixed states. The 
Hall resistivity $\rho_{xy}$ for both compounds is negative in the 
normal as well as in the mixed state and has no sign reversal below 
$T_c$ typical for high-$T_c$ superconductors.  $\em 
In~the~mixed~state$ the behavior of both systems is quite 
similar. The scaling relation $\rho_{xy}\sim\rho_{xx}^\beta$ 
($\rho_{xx}$ is the longitudinal resistivity) was found with 
$\beta=2.0$ and 2.1 for annealed Lu- and Y-based compounds, 
respectively.  The scaling exponent $\beta$ decreases with increasing
degree of disorder and can be varied by annealing.  This is 
attributed to a variation of the strength of flux pinning.  $\em 
In~the~normal~state$ weakly temperature dependent Hall coefficients 
were observed for both compounds. A distinct nonlinearity in the 
$\rho_{xy}$ dependence on field $H$ was found for LuNi$_2$B$_2$C in 
the normal state below 40~K, accompanied by a large magnetoresistance 
(MR) reaching +90\% for $H=160$~kOe at $T=20$~K. At the same time for 
YNi$_2$B$_2$C only linear $\rho_{xy}(H)$ dependences were observed in 
the normal state with an approximately three times lower MR value.  
This difference in the normal state behavior of the very similar Lu- 
and Y-based borocarbides seems to be connected with the difference in 
the topology of the Fermi surface of these compounds.

\end{abstract} 

\pacs{PACS numbers: 74.72.Ny; 72.15.Gd; 74.60.Ec}

\vskip2pc]
 
\narrowtext

\section{Introduction}
Investigation of the Hall effect in the normal and superconducting 
mixed states gives an important information about the electronic 
structure and the vortex dynamics of the investigated materials. The 
nature of both of them is not settled yet for the superconducting 
quaternary borocarbides $R\rm Ni_2B_2C$ ($R$=Y, rare 
earth).\cite{nagaraj,cava} Despite the fact that the borocarbides 
have a strongly anisotropic, layered tetragonal crystal structure, 
their electronic properties indicate three-dimensionality showing 
only a little anisotropy.  
\cite{lee,pickett,kim,johnston,fisher_prb97,metlushko,rath} 
Borocarbides based on magnetic rare earths show a wide range of 
competing effects between superconductivity and magnetism, see, e.g., 
Ref.\ \onlinecite{eisaki}.  One of the interesting features of some 
borocarbides ($R$=Er, Lu, Y) is the vortex lattice (VL) with unusual 
square symmetry 
\cite{dewilde,eskildsen,yethiraj_prl97,eskildsen_prlj97} observed in 
the mixed state for magnetic fields $H$ directed along tetragonal 
$c$-axis at $H \gtrsim 1$~kOe.  Square symmetry of VL can be 
connected \cite{dewilde,metlushko} with the anisotropy of the upper 
critical magnetic field $H_{c_2}(T)$ observed in the $ab$-plane for 
$\rm LuNi_2B_2C$.\cite{metlushko,rath} Practically no anisotropy of 
$H_{c_2}(T)$ was found for $\rm YNi_2B_2C$, \cite{johnston,rath} 
although this compound is very similar to $\rm LuNi_2B_2C$. The 
reason for the difference in the behavior of these two borocarbides 
is still unclear.

To our knowledge, no data on the Hall effect for $\rm LuNi_2B_2C$ and 
only few for some other borocarbides are known so 
far.\cite{fischer,narozh_jltp,mandal,narozh_jac} Namely, normal state 
Hall coefficients $R_H$ were found to be negative and only weakly 
temperature dependent for polycrystalline borocarbides based on $R$ = 
Y,\cite{fischer,narozh_jltp,mandal} Ho,\cite{fischer,mandal} 
La,\cite{fischer} and Gd \cite{mandal}. A negative but strongly 
temperature dependent Hall coefficient was found for the 
heavy-fermion-like compound $\rm YbNi_2B_2C$. \cite{narozh_jac} No 
sign reversal of the Hall resistivity $\rho_{xy}$ in the mixed state 
typical for high-$T_c$ superconductors was observed in $\rm 
YNi_2B_2C$,\cite{narozh_jltp} prepared under high pressure. The 
mixed-state Hall effect was not yet systematically studied for 
borocarbides. Since the mixed state Hall effect may depend on the 
peculiarities of the vortex lattice, it is of interest to investigate 
it for $\rm LuNi_2B_2C$ and $\rm YNi_2B_2C$ with an anomalous square 
VL. Also it is interesting to compare the results on the Hall effect 
in the normal and in the mixed states for Lu- and Y-based 
borocarbides, having substantially different types of anisotropy 
of the upper critical field. 

The Hall effect in the superconducting mixed state, which was studied 
up to now mainly for high-$T_c$ superconductors, has recently 
attracted a considerable attention and should be described in more 
detail. The magnetic field penetrates into a type-II superconductor 
by quantum vortices. In a transport current the flux lines experience 
the Lorentz force density:

\begin{equation} {\bf F}=\frac{1}{c}{\bf j}\times {\bf B}, 
\label{lorentz}
\end{equation} 

\noindent
where ${\bf j}$ is the transport current density and {\bf B} is the 
magnetic induction. The motion of vortex lines induces a
macroscopic electric field {\bf E} given by the 
relation:\cite{josephson}

\begin{equation} 
{\bf E}=-\frac{1}{c}{\bf v}_{L}\times {\bf B},
\label{josephson}
\end{equation} 

\noindent
where ${\bf v}_{L}$ is the velocity of vortex motion. The vortex 
motion along the Lorentz force (perpendicular to {\bf j}) gives the 
dissipative field $({\bf E\parallel j})$ and leads to the flux-flow 
resistivity. At the same time the vortex motion along the direction 
of transport current results in the Hall electric field $({\bf E\perp 
j,B)}$. Thus, the Hall effect is a sensitive test of vortex dynamics 
in the investigated material. On the other hand, normal carriers in 
the vortex core, experiencing a Lorentz force, can also give a 
contribution to the mixed state Hall effect by the usual mechanism. 

Two unexpected effects have been experimentally found for high-$T_c$ 
superconductors: (i) a sign reversal of the Hall resistivity 
$\rho_{xy}$ below $T_c$ and (ii) a striking scaling relationship 
between $\rho_{xy}$ and the longitudinal resistivity $\rho_{xx}$ in 
the superconducting transition region, $\rho_{xy}\sim 
\rho_{xx}^\beta$.

$\it{Sign~reversal}$ of the Hall resistivity $\rho_{xy}$ has been 
observed experimentally over a range of temperatures and magnetic 
fields below $T_c$ for several types of high-$T_c$ superconductors, 
e.g., $\rm YBa_2Cu_3O_{7-y}$,\cite{galffy,ri} $\rm
Bi_2Sr_2CaCu_2O_8$,\cite{ri,samoilov_93} $\rm 
Tl_2Ba_2CaCu_2O_8$,\cite{budhani} $L\rm_{2-x}Ce_xCuO_4$ ($L$ = Nd, 
Sm),\cite{hagen_prb93,cagigal} $\rm YBa_2Cu_3O_{7-y}$/ $\rm
PrBa_2Cu_3O_{7-y}$ superlattices \cite{li,wang_prl97,yang_pr97} 
(YBCO, BSCCO, TBCCO, LCCO, YBCO/ PBCO respectively), as well as for 
some conventional superconductors:  In-Pb alloys, V, Nb (see 
Ref.\ \onlinecite{parks}), $\rm Mo_3Si$,\cite{smith} $2H\rm-
NbSe_2$.\cite{bhat} This Hall effect anomaly cannot be understood 
within the framework of the classical Bardeen-Stephen\cite{BS} and 
Nozi\`eres-Vinen\cite{NV} theories of vortex motion predicting the 
same sign of the Hall voltage for both the superconducting and the 
normal state.  Recently several models based on different approaches 
have been proposed for the description of this effect (see, e.g., 
Refs.\ 
\onlinecite{wang_prl91,dorsey_pr92,kopnin,aronov,otterlo,khomskii} 
and references therein), but the origin of this phenomenon remains a 
controversial problem. Meanwhile, the sign reversal of $\rho_{xy}$ 
below $T_c$ is expected to be not a universal property, but it's 
existence seems to dependent crucially on the peculiarities of the 
electronic structure.\cite{dorsey_pr92,kopnin,aronov,otterlo} 
Experimentally the pronounced influence of the doping level on the 
sign of the Hall voltage close to $T_c$ was observed for various 
high-$T_c$ cuprates. \cite{jin_98,nagaoka_98} The sign reversal of 
the Hall effect disappears for heavily underdoped \cite{jin_98} and 
strongly overdoped \cite{nagaoka_98} regimes.
                                 
$\it{Scaling~behavior}$, $\rho_{xy}$$\sim$$\rho_{xx}^\beta$, in the 
superconducting mixed state was observed for the first time by 
Luo~{\it et al.} \cite{luo} for an YBCO thin film ($\beta=1.7$). The 
same relationship was also found for several types of high-$T_c$ 
superconductors:  YBCO single crystals ($\beta\approx 1.7$), 
\cite{rice} BSCCO ($\beta\approx 2$), \cite{samoilov_93} TBCCO 
($\beta\approx 2$), \cite{budhani,samoilov_pr94} LCCO ($\beta\approx 
0.8$), \cite{cagigal} (YBCO/PBCO) superlattices ($\beta\approx 1.7$).  
\cite{yang_pr97,li} In a recent investigation of superconducting 
indium thin films scaling with $\beta$ value $2\div 3$ was observed.  
\cite{okuma} Theoretically, Dorsey and Fisher \cite{dorsey_fischer} 
(DF) have interpreted the observed behavior in the framework of 
glassy scaling near a vortex-glass transition. In their model, 
assuming the existence of a vortex-glass transition in a 
three-dimensional vortex system, the region where scaling behavior 
should be observed is restricted to a narrow region near the 
vortex-glass transition. However, it should be mentioned that scaling 
behavior was observed far beyond the possible vortex-glass 
transition. \cite{ri} A phenomenological model, based on an entirely 
different approach, has been proposed by Vinokur~{\it et al.}, 
\cite{vinokur_prl93} who have calculated the effect of pinning on the 
Hall resistivity. In their model the Hall conductivity 
$\sigma_{xy}\cong\rho_{xy}/\rho_{xx}^2$ ($|\rho_{xy}|\ll\rho_{xx}$) 
is independent of disorder and the scaling law 
$\rho_{xy}$$\sim$$\rho_{xx}^2$ is believed to be a general feature of 
any vortex state with disorder-dominated dynamics.  Therefore, the 
value of $\beta=2$ should not depend on the degree of disorder. On 
the other hand, Wang, Dong and Ting \cite{wang_prl94} (WDT) recently 
modified their earlier work, \cite{wang_prl91} based on the normal 
core model proposed by Bardeen and Stephen. \cite{BS} They developed 
a theory for the Hall effect including both pinning and thermal 
fluctuations. In the WDT theory scaling $\em and \/$ sign reversal of 
$\rho_{xy}$ are explained by specially taking into account the 
backflow current of vortices due to pinning.  \cite{wang_prl94} 
Thereby, $\beta$ changes from 2 to 1.5 as the pinning strength 
increases. \cite{wang_prl91,wang_prl94} Controversial experimental 
results have been reported on the influence of disorder on the 
mixed-state Hall effect. For irradiated YBCO samples, $\beta$ was 
found to be $1.5\pm 0.1$ compared to $2\pm 0.2$ for unirradiated 
ones, \cite{kang} in accordance with WDT (see also Refs.\ 
\onlinecite{li,wang_prl97,yang_pr97}).  However, no influence of 
disorder on the scaling exponent was observed for TBCCO irradiated by 
heavy ions. In that case, $\beta=1.85$ holds even after irradiation, 
\cite{budhani} (see also Ref.\ \onlinecite{samoilov_prl95}). A strong 
influence of pinning on the Hall effect in the mixed state was 
observed of YBCO single crystals.  \cite{morgoon} At the same time it 
was pointed out in Ref.\ \onlinecite{jin_98} that pinning effects 
cannot be the only reason for the Hall anomaly for YBCO single 
crystals.  All these controversial results show that more work is 
necessary for better understanding of the mixed-state Hall effect and 
the influence of disorder on it.
 
Very recently Wang and Maki \cite{maki_prb98} have interpreted the 
anisotropy of $H_{c_2}(T)$ observed for borocarbides in terms of a 
three-dimensional version of $d_{x^2-y^2}$ superconductivity. 
Possible $d$ wave nature of superconductivity for borocarbides gives 
an additional motivation for further study of their electronic 
properties. In the present study we have investigated the Hall effect 
in the normal as well as in the mixed state for $\rm LuNi_2B_2C$ and 
$\rm YNi_2B_2C$ compounds prepared under the same conditions. The 
results for $\rm LuNi_2B_2C$ have been briefly reported in 
Ref.\ \onlinecite{narozh_ssc}.  
 
\section{Experimental details} Polycrystalline $\rm LuNi_2B_2C$ (in 
the following denoted as PC AN) and $\rm YNi_2B_2C$ samples were 
prepared by arc-melting in Ar atmosphere and subsequent careful 
annealing at 1100$\rm\ ^\circ C$, as described in more detail in 
Ref.\ \onlinecite{muller_96}.  The phase purity of the samples was 
checked by X-ray diffraction  on a Philips PW 1820 system with $\rm 
CoK_\alpha$ radiation.  The reflections revealed practically a single 
phase. The lattice parameters were $a$=3.464~{\AA}, $c$=10.635~{\AA} 
for $\rm LuNi_2B_2C$ and $a$=3.528~{\AA}, $c$=10.546~{\AA} for $\rm 
YNi_2B_2C$. Bar-shaped samples were cut from the ingots. Typical 
dimensions of the samples were $\rm 3\times 1\times 0.3~mm^3$. Hall 
contacts with typical misalignment of less than 0.1 mm were used 
(this is essential because the maximum of the Hall voltage does not 
exceed several tens of nanovolts).  At each point the Hall voltage 
was measured for two inverse directions of the magnetic field.  Most 
measurements of the Hall effect and of the $\em ac$-susceptibility 
were done in magnetic fields up to 50~kOe using a Lake Shore model 
7225 susceptometer with Keithley 182 nanovoltmeter and PAR-5209 $\em 
Lock-in\/$ amplifier.  Some measurements in magnetic fields up to 160 
kOe were performed using an Oxford Teslatron system.  The values of 
electrical current were 10-20 mA for $\em dc$ measurements and 1 mA 
for $\em ac$ ones.  The magnetoresistance (MR) was measured by the 
standard four-probe method. For comparison some measurements were 
performed on an unannealed $\rm LuNi_2B_2C$ sample (denoted as PC 
UNAN) wich has a considerably higher degree of disorder.  

\section{Results and discussion} 
\subsection{Resistivity and upper critical magnetic field} 
The experiments in the present work have been performed on 
polycrystalline samples (to the best of our knowledge, no data on the 
Hall-effect for single crystalline borocarbides have been reported so
far). For characterization of our samples the results on 
resistivity, upper critical field and magnetoresistance for them will 
be compared with data known for single crystals.

\begin{figure} 
\epsfxsize=7.4cm 
\centerline{\epsfbox{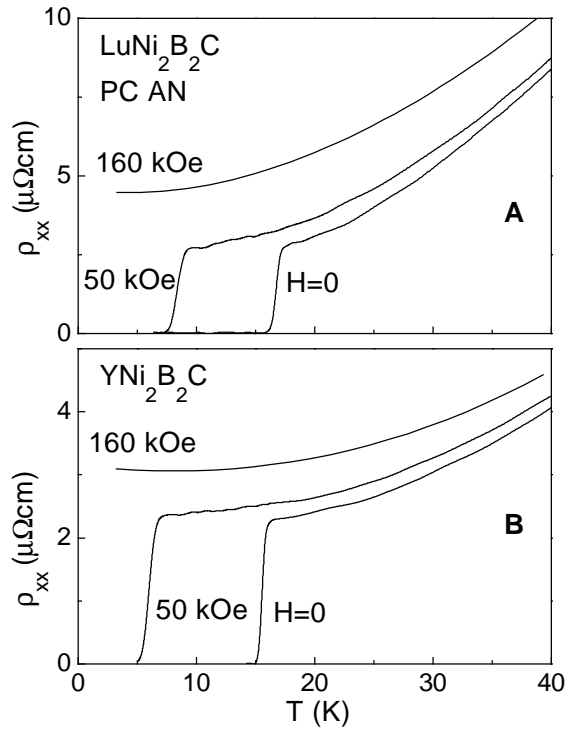}} 
\vspace{0.5pc} 
\caption{Longitudinal resistivity $\rho_{xx}$ as a function of 
temperature $T$ at $H$=0, 50, and 160~kOe for the annealed $\rm 
LuNi_2B_2C$ and $\rm YNi_2B_2C$ samples.} 
\label{rho_xx(T)} 
\end{figure}

The temperature dependencies of the longitudinal resistivity 
$\rho_{xx}(T)$ for the annealed $\rm LuNi_2B_2C$ and $\rm YNi_2B_2C$ 
samples are depicted in Fig.\ \ref{rho_xx(T)}. The $\rho_{xx}(T)$ 
curves obtained at $H$=50~kOe and 160~kOe are also shown. The 
resistivity of both compounds exhibits a weak temperature dependence 
just above superconducting transition temperature $T_c$. Both samples 
have a rather sharp superconducting transition, a low resistivity at 
low temperatures and high values of $T_c$ and the residual resistance 
ratio RRR (RRR=$\rho_{xx}$(300~K)/$\rho_{xx}$(17~K)). These 
parameters are compared in Table\ \ref{table1} with those reported
recently for $\rm LuNi_2B_2C$ and $\rm YNi_2B_2C$ single crystals.  
\cite{shulga,rath,fisher_prb97,metlushko} It should be pointed out, 
that the resistivity of borocarbides is practically isotropic.  The 
small difference ($\approx 2\% $) between the in-plane resistivity 
$\rho_a$ and the resistivity along the $c$-axis $\rho_c$ observed for 
$\rm YNi_2B_2C$ single crystals \cite{fisher_prb97} at $T$=15$\div 
$300~K is well within the experimental uncertainty. Thus, it is 
reasonable to compare the values of resistivity for polycrystalline 
and single crystalline borocarbides.

\begin{figure} 
\epsfxsize=6.9cm 
\centerline{\epsfbox{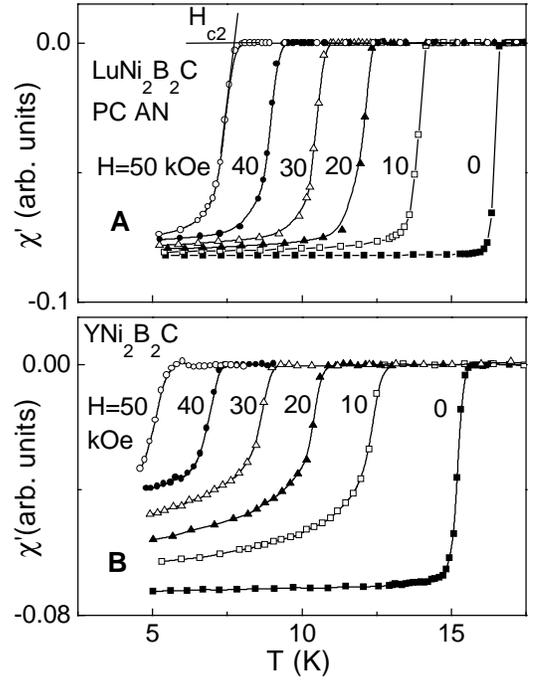}} 
\vspace{0.5pc} 
\caption{Temperature dependence of the real part of the $ac$-magnetic 
susceptibility for the $\rm LuNi_2B_2C$ (A) and $\rm YNi_2B_2C$ (B) 
samples in several magnetic fields. The upper critical field $H_{c2}$ 
was determined by linear extrapolation of the $\em 
ac\/$-susceptibility curve to zero susceptibility  value, as shown in 
the upper part of the figure.  Lines are guides for the eye.} 
\label{chi(T)} 
\end{figure}

More precisely, the value of $\rho_{xx}$(17~K) for the annealed 
polycrystalline $\rm LuNi_2B_2C$ sample is 2.7~$\mu\Omega$cm, which 
is close to that for $\rm LuNi_2B_2C$ single crystals 
(1.6$\div$2.5~$\mu\Omega$cm, Refs.\ 
\onlinecite{shulga,rath,fisher_prb97,metlushko}), see Table\ 
\ref{table1}.  The value of RRR is 41 for our $\rm LuNi_2B_2C$ 
sample, which is $\em significantly~higher \/$ than those observed 
for single crystals (23$\div$27, see Refs.\ 
\onlinecite{shulga,rath,fisher_prb97}). Also the value of 
$T_c$=16.7~K is slightly higher than that reported for single 
crystals (15.8$\div$16.5~K, see Refs.\ 
\onlinecite{shulga,rath,fisher_prb97,metlushko}). The width of the 
superconductig transition $\Delta T_c$ (determined from zero field 
$\em ac \/$-susceptibility curve, see Fig.\ \ref{chi(T)}) is 0.27~K 
which is close to the values 0.2$\div$0.25~K typical for single 
crystals. \cite{shulga,rath,fisher_prb97}  The PC UNAN $\rm 
LuNi_2B_2C$ sample has a lower $T_c$ (14.7~K), a wider 
superconducting transition and a more than one order of magnitude 
higher value of $\rho_{xx}(17~K)$. 

For the polycrystalline $\rm YNi_2B_2C$ sample the values of 
$\rho_{xx}$(17~K), RRR, $T_c$ and $\Delta T_c$ are also comparable 
with the results reported for $\rm YNi_2B_2C$ single crystals, 
\cite{shulga,rath,fisher_prb97} see Table\ \ref{table1}. At the same 
time our $\rm YNi_2B_2C$ sample has an approximately two times lower 
RRR value than  $\rm LuNi_2B_2C$ prepared under the same conditions.

\begin{figure} 
\epsfxsize=7.4cm 
\centerline{\epsfbox{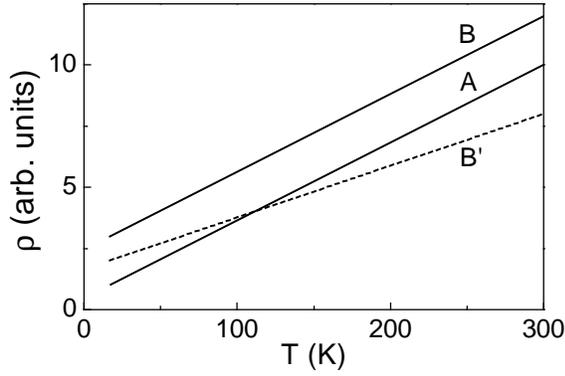}} 
\vspace{0.5pc} 
\caption{Temperature dependence of the resistivity for two 
imaginary samples (A and B) with different RRR values (10 and 4, 
respectively). Line B$'$ corresponds to the underestimated by 1/3 
of $\rho $ for sample B and has the same value of RRR as line 
B. See text for details.} 
\label{fig_add} 
\end{figure}

The results for the resistivity $\rho_{xx}$(300~K) collected in 
Table\ \ref{table1} show surprisingly large discrepancies even for 
the single crystals. Thus an intersection of the $\rho_{xx}(T)$ 
dependences can be recognised for the two $\rm YNi_2B_2C$ single 
crystals. \cite{shulga,rath} These facts could be naturally 
understood taking into account the large uncertainty in geometrical 
factor used to determine the value of $\rho_{xx}$ from the 
experimentally measured resistance especially for small single 
crystals. To clarify this, linear $\rho_{xx}(T)$ dependences for two 
imaginary samples of the same imaginary compound with different RRR 
values (10 and 4 for samples A and B, respectively) have been plotted 
in Fig.\ \ref{fig_add}.  Mattissen's rule is expected to be valid for 
them, i.e. the difference in resistivities does not dependent on 
temperature.  Let us suppose, that the measurements of $\rho_{xx}(T)$ 
give the "true" values for the sample A and underestimate it by 1/3 
of its "true" value for the sample B (e.g., due to the uncertainty in 
the dimensions of the sample, final width of the contacts, etc.). In 
that case for the sample B the obtained ("measured") $\rho_{xx}(T)$ 
curve (denoted as B$'$ on  Fig.\ \ref{fig_add} and having the same 
value of RRR as the curve B) will cross the curve for the sample A.  
This example illustrates that an uncertainty of geometrical factor of 
$\approx 20\div 30\%$ could explain the difference in room 
temperature resistivities and surprising intersection of 
the $\rho_{xx}(T)$ dependences for the two $\rm YNi_2B_2C$ single 
crystals.  \cite{shulga,rath} A strong support for this
explanation is that, for single crystals of 
different quality known, one would expect close resistivity values 
rather at room temperature than at low temperatures. We conclude that 
the quality of the borocarbide samples compared in Table\ 
\ref{table1} can be mainly judged from the RRR data, whereas 
resistivity values are strongly influenced by the uncertainty in the 
geometrical factor used for determination of $\rho_{xx}$. Noteworthy, 
both room and low temperature resistivities for our $\rm 
LuNi_2B_2C$ sample are close to those reported for the $\rm 
YNi_2B_2C$ single crystal wich has a similar value of RRR 
\cite{shulga} as our policrystalline $\rm LuNi_2B_2C$ sample (see 
Table\ \ref{table1}).
  
\begin{figure} 
\epsfxsize=7.4cm 
\centerline{\epsfbox{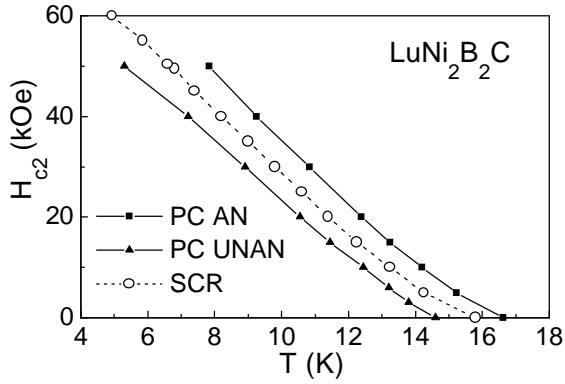}} 
\vspace{0.5pc} 
\caption{Temperature dependence of the upper critical field 
$H_{c_2}$ for three $\rm LuNi_2B_2C$ samples. Open symbols -
the results of Ref.\ \protect \onlinecite{metlushko} for single 
crystal.  Solid symbols - the results for PC AN and PC UNAN 
polycrystalline samples (see text).  Lines are guides for the eye.} 
\label{Hc2(T)} 
\end{figure}

Superconducting transitions determined from $\em ac 
\/$-susceptibility measurements are shown in Fig.\ \ref{chi(T)} 
for different magnetic fields. Temperature dependences of upper 
critical magnetic fields $H_{c_2}(T)$ are depicted on Fig.\ 
\ref{Hc2(T)} for the annealed and unannealed $\rm LuNi_2B_2C$ 
samples. (The value of $H_{c_2}$ was determined, similar as in
Ref.\ \onlinecite{metlushko}, by the extrapolation of the 
$\em{ac}$-susceptibility curve to zero susceptibility value, see 
Fig.\ \ref{chi(T)}A.) For comparison the data from Ref.\ 
\onlinecite{metlushko} for a $\rm LuNi_2B_2C$ single crystal (SCR) 
with $H\parallel$$<$110$>$ are also shown. The upward curvature (UC) 
in the $H_{c_2}(T)$ dependence is clearly visible near $T_c$. Note 
that, in accordance with Ref.\ \onlinecite{shulga}, the UC region is 
more pronounced and $|\partial H_{c_2}/\partial T|$ is higher for the 
annealed PC AN sample. This suggests, \cite{shulga} that this sample 
is close to the clean limit in terms of traditional theory of type-II 
superconductors. It is of interest to compare $|\partial 
H_{c_2}/\partial T|$ values for different samples determined from the 
approximately linear parts of the $H_{c_2}(T)$ dependences 
($H$=20$\div$50~kOe).  The value of $|\partial H_{c_2}/\partial 
T|$=6.8~kOe/K, obtained for the annealed $\rm LuNi_2B_2C$ sample, is 
in good agreement with those determined for single crystals:  
6.7~kOe/K (calculated by arithmetic averaging of $|\partial 
H_{c_2}/\partial T|$ data reported in Ref.\ \onlinecite{metlushko} for 
three directions of $H$, $\parallel<$100$>$,$<$110$>$, and 
$<$001$>$), 6.4~kOe/K (reported for $H\parallel$$<$001$>$ for another 
single crystal in Ref.\ \onlinecite{shulga}) and 6.2~kOe/K (calculated by 
arithmetic averaging of the data reported in Ref.\ \onlinecite{rath} for 
$H\parallel<$100$>$ and $<$001$>$). The $H_{c_2}(T)$ dependence for 
our $\rm YNi_2B_2C$ sample is similar to that observed for the $\rm 
LuNi_2B_2C$ PC AN one and also is in good agreement with the results 
reported for $\rm YNi_2B_2C$ single crystals. \cite{shulga,rath} The 
value of $|\partial H_{c_2}/\partial T|$ for our $\rm YNi_2B_2C$ 
sample is 6.0~kOe/K, see Table\ \ref{table1}.

The low resistivity just above $T_c$, high RRR and $T_c$ values,
narrow superconducting transitions, pronounced UC region in the 
$H_{c_2}(T)$ dependences and X-ray diffraction results give evidence 
for a good quality of our annealed $\rm LuNi_2B_2C$ and $\rm 
YNi_2B_2C$ samples. 

\begin{figure} 
\epsfxsize=8.3cm 
\centerline{\epsfbox{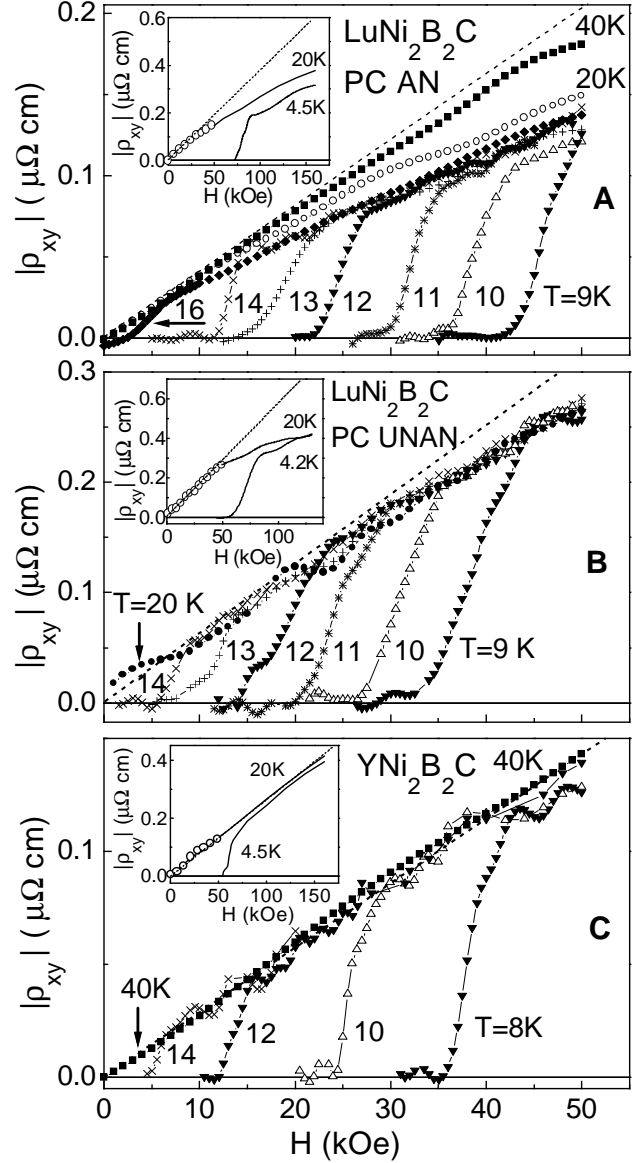}} 
\vspace{0.5pc} 
\caption{Absolute value of the Hall resistivity $|\rho_{xy}|$ as a 
function of magnetic field $H$ for the annealed (A) and unannealed 
(B) $\rm LuNi_2B_2C$ as well as for $\rm YNi_2B_2C$ (C)  
samples. The dashed lines are low-field asymptotes to 
the normal state curves. The inserts show the results for $H$ up to 
160~kOe.  Open circles in the insets denote the data obtained
for $H\leq$ 50~kOe. Only some representative curves and data points 
are shown.} 
\label{rho_xy(H)} 
\end{figure}

\subsection{Normal state Hall effect} The temperature dependences of
the Hall resistivity, $\rho_{xy}(H)$, for $\rm LuNi_2B_2C$ PC~AN and 
PC UNAN samples as well as for $\rm YNi_2B_2C$ sample in the normal 
and in the mixed states are shown in Fig.\ \ref{rho_xy(H)}. First of 
all it should be emphasized that the Hall resistivity of all the 
samples is negative at 3.3~K$\leq T\leq$300~K, and has no sign 
reversal below $T_c$. 

$\it{In~the~normal~state}$, a pronounced nonlinearity in the 
$\rho_{xy}(H)$ dependences is evident at $T\lesssim$ 40~K for both 
$\rm LuNi_2B_2C$ samples. Linear $\rho_{xy}(H)$ dependences 
extrapolated from the low fields region are also shown in Fig.\ 
\ref{rho_xy(H)} by dashed lines. The deviation from linear 
$\rho_{xy}(H)$ dependence increases with lowering temperature. The 
anomaly is more pronounced for the annealed sample, although it is 
also distinctly seen for the unannealed one.  More clearly the 
nonlinearity in the $\rho_{xy}(H)$ dependences can be seen in the 
insets of Fig.\ \ref{rho_xy(H)}A and B where some results obtained in 
high magnetic fields (up to 160~kOe) are presented. It should be 
underlined that no nonlinearity in the $\rho_{xy}(H)$ dependence was 
observed for our $\rm YNi_2B_2C$ sample (see the inset of Fig.\ 
\ref{rho_xy(H)}C).  Earlier, linear $\rho_{xy}(H)$ dependences have 
been reported for $\rm YNi_2B_2C$ \cite{narozh_jltp} and $\rm 
YbNi_2B_2C$ \cite{narozh_jac} samples prepared under high pressure. 
No indications of nonlinear $\rho_{xy}(H)$ dependences have been 
observed for Y-, Ho-, Gd- and La-based borocarbides. 
\cite{fischer,mandal} Thus, the nonlinearity in the $\rho_{xy}(H)$ 
dependence, found for $\rm LuNi_2B_2C$ samples with essentially 
different quality, can be considered as an intrinsic and specific 
property of the Lu-based borocarbide.
 
A nonlinear and even nonmonotonous $\rho_{xy}(H)$ dependence has 
been found earlier for the heavy fermion superconductor $\rm 
UBe_{13}$ \cite{aleks} and have been interpreted in the framework of 
a two-band model. \cite{aleks} In this model, at low fields, the 
light carriers with high mobilities give the prevalent contribution 
to the Hall effect, whereas at high fields the contribution of the 
heavier carriers having lower mobilities is more significant. Very 
recently a similar two-band model \cite{gollnik_prb98} has been used 
to interpret the results on the transport properties of $\rm 
Nd_{2-x}Ce_xCuO_4$ epitaxial thin films. In an entirely different 
type of multi-band models, \cite{shulga} the existence of at least 
two bands with significantly different Fermi-velocities was found to 
be very essential for the quantitative description of $H_{c_2}(T)$ 
curves with sizable UC for the $\rm Lu$- and $\rm Y$-based 
borocarbides. Several groups of carriers with different effective
masses have been directly observed for $\rm YNi_2B_2C$ in dHvA 
experiments. \cite{nguyen} Thus, some kind of a two-band model may be 
applicable for understanding of the nonlinear $\rho_{xy}(H)$ 
dependence found by us for $\rm LuNi_2B_2C$ borocarbide.

\begin{figure} 
\epsfxsize=7.5cm 
\centerline{\epsfbox{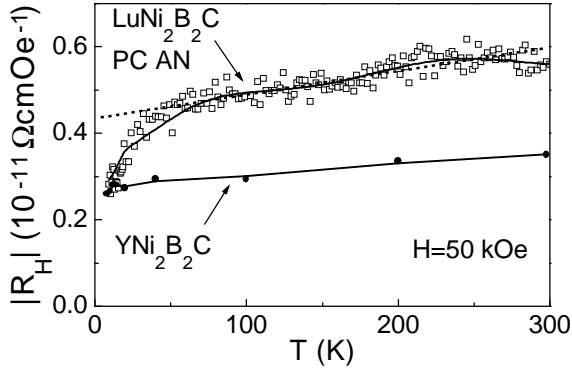}} 
\vspace{0.5pc} 
\caption{Absolute value of the Hall coefficient $|R_H|$  (obtained at 
$H$=50~kOe) as a function of temperature $T$ for $\rm LuNi_2B_2C$ and 
$\rm YNi_2B_2C$ samples.  Dotted line - linear extrapolation of high 
temperature data for $\rm LuNi_2B_2C$. Solid lines are guides for the 
eye.} 
\label{R_H(T)} 
\end{figure} 

In Fig.\ \ref{R_H(T)} the temperature dependences of the Hall 
coefficients $R_H(T)=\rho_{xy}(T,H)/H$ at $H$=50~kOe are shown for 
the annealed $\rm LuNi_2B_2C$ and $\rm YNi_2B_2C$ samples. Below
$\sim$60~K the $R_H(T)$ dependence for $\rm LuNi_2B_2C$ shows a 
considerable deviation from the dotted line describing $R_H(T)$ at 
$H$=50~kOe for higher temperatures. This deviation is obviously 
connected with the nonlinearity in $\rho_{xy}(H)$ curves shown in 
Fig.\ \ref{rho_xy(H)}A. At low temperatures the values of the Hall 
coefficient on the dotted line in Fig.\ \ref{R_H(T)} (obtained
by the extrapolation of the high temperature $R_H(T)$ curve) coincide 
with the values of the low-field Hall coefficient, calculated at low 
temperatures using the low-field asymptotes for the $\rho_{xy}(H)$ 
curves (shown in Fig.\ \ref{rho_xy(H)}A by dashed line). Only weak 
temperature dependences were observed for the low field Hall 
coefficients of $\rm LuNi_2B_2C$ and $\rm YNi_2B_2C$ borocarbides.  
This is in agreement with the observation of a weak $R_H(T)$ 
dependence for $\rm YNi_2B_2C$ in Refs.\ 
\onlinecite{fischer,narozh_jltp,mandal}. Weak $R_H(T)$ dependences 
were reported also for La-, Ho- and Gd-based borocarbides.  
\cite{fischer,mandal} Below $\sim$60~K, the $R_H(T)$ curve obtained 
for $\rm LuNi_2B_2C$ at $H$=50~kOe exhibits a pronounced temperature 
dependence connected with the nonlinearity found for 
$\rho_{xy}(H)$ at low temperatures. Noteworthy, that a strong 
decrease of the Hall coefficient was found with increasing
temperature for $\rm YbNi_2B_2C$ \cite{narozh_jac} borocarbide having 
moderate heavy-fermion-like behavior. The values of $R_H$ obtained in 
this work for $\rm LuNi_2B_2C$ and $\rm YNi_2B_2C$ are comparable 
with those earlier reported for $\rm YNi_2B_2C$, 
\cite{fischer,mandal} but they are about five times ($\rm 
LuNi_2B_2C$) or ten times ($\rm YNi_2B_2C$) smaller than the value 
resulting from band structure calculations \cite{pickett} for $\rm 
LuNi_2B_2C$ ($3\cdot 10^{-9}$m$^3$/C=3$\cdot 10^{-11}\Omega$cm/Oe).  
These deviations may be caused by correlation effects in 
borocarbides. The estimation of the carrier density from the $R_H$ 
value at $T$=300~K, by using a single band model which is a rough 
approximation, gives 1.5 and 2.4 carriers per unit cell for Lu- and 
Y-based borocarbides, respectively. (The estimation of the carrier 
density for $\rm YNi_2B_2C$ prepared under high 
pressure gives 0.6 carriers per unit cell \cite{narozh_jltp}, 
i.e. about four times smaller than present result and values reported 
in Refs.\onlinecite{fischer,mandal}. Probably this difference is 
connected with high sensitivity of the Hall coefficient for 
method of sample preparation.)

\begin{figure} 
\epsfxsize=8.3cm 
\centerline{\epsfbox{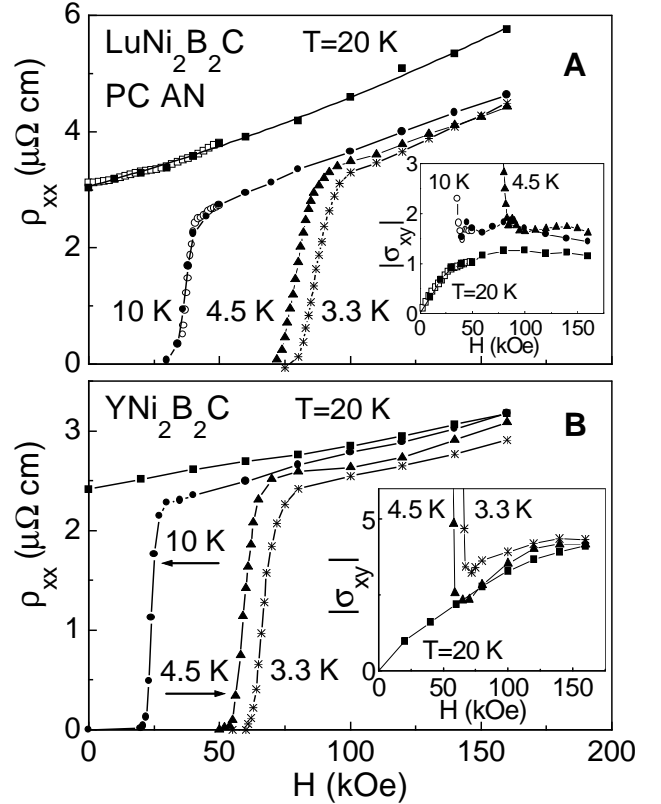}} 
\vspace{0.5pc} 
\caption{Magnetic-field dependence of the longitudinal resistivity
$\rho_{xx}$ for $\rm LuNi_2B_2C$ and $\rm YNi_2B_2C$. In the insets 
absolute value of the Hall conductivity $|\sigma_{xy}|$ (in 
$10^{-2}\mu\Omega^{-1}cm^{-1}$) vs. field is shown.  Open
circles in part~A denote the results obtained for $H\leq$
50~kOe.  Lines are guides for the eye.} 
\label{rho_xx(H)} 
\end{figure} 
 
\subsection{Normal state magnetoresistance}
In Fig.\ \ref{rho_xx(H)} the field dependence of the longitudinal 
resistivity $\rho_{xx}(H)$ is shown for the annealed $\rm LuNi_2B_2C$ 
and $\rm LuNi_2B_2C$ samples. The values of magnetoresistance 

\begin{equation} 
{\rm MR}=\frac{\rho_{xx}(H)-\rho_{xx}(0)}{\rho_{xx}(0)}
\label{MR} 
\end{equation} 

\noindent
for the annealed $\rm LuNi_2B_2C$ sample at $T=20$~K are as high as 
25\% and 90\% for $H=50$~kOe and 160~kOe, respectively (see also 
Fig.\ \ref{rho_xx(T)}). At the same time considerably smaller values 
of MR were observed at these fields (10\% and 33\%, respectively) for 
the $\rm YNi_2B_2C$ sample, prepared under the same conditions as the
Lu-based one, see Table\ \ref{table1}. It should be underlined, that 
a magnetoresistance  of only $\approx$7.3\% was observed, at 
$H$=45~kOe and $T=20$~K, for a $\rm LuNi_2B_2C$ single crystal with 
RRR=25 ($H$ parallel to the tetragonal $c$-axis), \cite{rath} i.e.  
the value of MR for our $\rm LuNi_2B_2C$ polycrystalline sample is 
about 3.5 times higher, than that of this single crystal. The value
of MR (7.5\%) reported in Ref.\ \onlinecite{rath} for $\rm YNi_2B_2C$ 
single crystal with RRR=18 ($H\parallel c$) is comparable with that 
found for investigated $\rm YNi_2B_2C$ sample. High values of MR
can be considered as an additional indication of the high quality of 
our annealed samples, because the value of MR for the unannealed 
sample is approximately 40 times smaller (see Table\ \ref{table1}).
It should be also noted that the impurities of magnetic rare earths, 
the concentration of which could be greater in Lu-based sample than 
in Y-based one, can not lead to the increase of MR, because the 
introduction of magnetic ions to nonmagnetic substance (e.g., 
borocarbide lattice) gives the negative contribution to MR due to 
decrease of spin-disorder scattering in a magnetic field, see, e.g., 
Refs. \onlinecite{fisher_prb97,narozh_jltp}.

A possible reason for the very large positive MR in $\rm LuNi_2B_2C$ 
and for the significantly larger MR of the polycrystalline sample 
compared to the single crystal in Ref.\ \onlinecite{rath} is the 
formation of open orbits on the Fermi surface of that compound for 
$H$$\perp$$c$.  (In principle the possibility of the formation of 
open orbits for borocarbides was pointed out in band structure 
calculations. \cite{kim,lee} In Ref.\ \onlinecite{kim}, e.g., it was 
claimed that one part of the Fermi surface is a cylinder along the 
$c$-axis.) It is well known \cite{lifshitz} that open orbits can lead 
to large values of MR~$\propto H^2$, whereas closed orbits should 
give rise to $\it {saturation}$ of magnetoresistance for large $H$.  
In that case, for polycrystals, the averaging of MR should lead to a 
practically linear $\rho_{xx}(H)$ dependence \cite{lifshitz} (so 
called Kapitza's law). In accordance with this, the observed 
$\rho_{xx}(H)$ dependences for our $\rm LuNi_2B_2C$ PC AN sample 
follow approximately a linear law, see Fig.\ \ref{rho_xx(H)}. The 
MR($H$) dependence for polycrystals, in the case of open orbits for 
some directions of $H$, should be $\em stronger \/$ than that 
observed for single crystals for $H$$\parallel$$c$ where only closed
orbits could be expected.  Therefore, the significantly larger MR 
found for the $\rm LuNi_2B_2C$ polycrystals, in comparison with that 
observed for the single crystal for $H$$\parallel$$c$, can be 
considered as an indication for the open-orbits formation in $\rm
LuNi_2B_2C$ for $H$$\perp$$c$. Investigation of the MR in high fields 
for $\rm LuNi_2B_2C$ single crystals with the two configurations (i) 
$\bf j$~$\parallel c$ and $H\perp c$ and (ii) $\bf j$~$\perp c$ and 
$H\parallel c$ are necessary to check this conclusion.

The nonlinear $\rho_{xy}(H)$ dependence and the large MR, found in 
this study, as well as the anisotropy of $H_{c_2}$ in the $ab$-plane, 
\cite{metlushko,rath} earlier reported for $\rm LuNi_2B_2C$, may be 
caused by the same reason, namely, by peculiarities of its electronic 
structure. It should be underlined, that  all these anomalies are 
absent for $\rm YNi_2B_2C$. (For $\rm YNi_2B_2C$ a linear 
$\rho_{xy}(H)$ dependence and a substantially smaller MR can be seen 
in Fig.\ \ref{rho_xx(H)} and only a very small anisotropy of 
$H_{c_2}(T)$ were reported in Refs.\ \onlinecite{johnston,rath}.) The 
differences in the properties of these very similar compounds should 
be connected with difference between their electronic structure.  As 
has been noted in Ref.\ \onlinecite{lee} the Fermi surface topology 
of the borocarbides is very sensitive to the position of the Fermi 
level, which may be slightly different for the two cases, Lu and Y, 
due to, e.g., different lattice constants. From the obtained results 
the formation of open orbits seems to be easier in case of $\rm
LuNi_2B_2C$ in comparison with $\rm YNi_2B_2C$. Nevertheless only the 
comparative study of $\rm LuNi_2B_2C$ and $\rm YNi_2B_2C$ single 
crystals (e.g., investigation of the angular dependence of MR in high 
fields) can give definitive verification of the proposed model.

Theoretically, it is more convenient to describe the behavior of the 
Hall effect in terms of the conductivity tensor rather than by the 
resistivity one, see, e.g., Ref.\ \onlinecite{lifshitz}. As shown in 
the inset of Fig.\ \ref{rho_xx(H)}A, the nonlinearity in the 
dependence on magnetic field of the Hall conductivity, 
$\sigma_{xy}(H)$, in the normal state for $\rm LuNi_2B_2C$ is even 
more pronounced than the nonlinearity in the Hall resistivity curve 
$\rho_{xy}(H)$ ($\sigma_{xy}\cong\rho_{xy}/\rho_{xx}^2$, 
$\rho_{xx}>>|\rho_{xy}|$).  It is interesting to note that 
$\sigma_{xy}$ for $\rm LuNi_2B_2C$ becomes practically independent of 
the magnetic field for $H=80\div$160~kOe, at $T=4.5\div$20~K (see 
Fig.\ \ref{rho_xx(H)}A).  The nonlinear $\rho_{xy}(H)$ dependence and 
the large MR of $\rm LuNi_2B_2C$ are probably closely connected and 
result in a practically constant $\sigma_{xy}(H)$ at high magnetic 
fields. The reason why $\sigma_{xy}$ is independent of $H$ for high 
fields, resulting in $\rho_{xy}\sim\rho_{xx}^2$ in the 
$\it{normal~state}$, is not yet understood. (It is noteworthy that 
$\rho_{xy}\sim \rho_{xx}^2$ scaling in the $\it{normal~state}$ state 
was earlier observed for the superconducting heavy fermion compound 
$\rm UBe_{13}$.  \cite{aleks}) At the same time the Hall conductivity 
of $\rm YNi_2B_2C$ has only a slight nonlinearity at 
$T$=20~K (see the inset of Fig.\ \ref{rho_xx(H)}B). Only at $T$=4.5~K 
and 3.3~K some tendency for saturation in $\sigma_{xy}(H)$ 
dependences was observed in high fields.

\subsection{Mixed state Hall effect} 
\begin{figure} 
\epsfxsize=7.6cm 
\centerline{\epsfbox{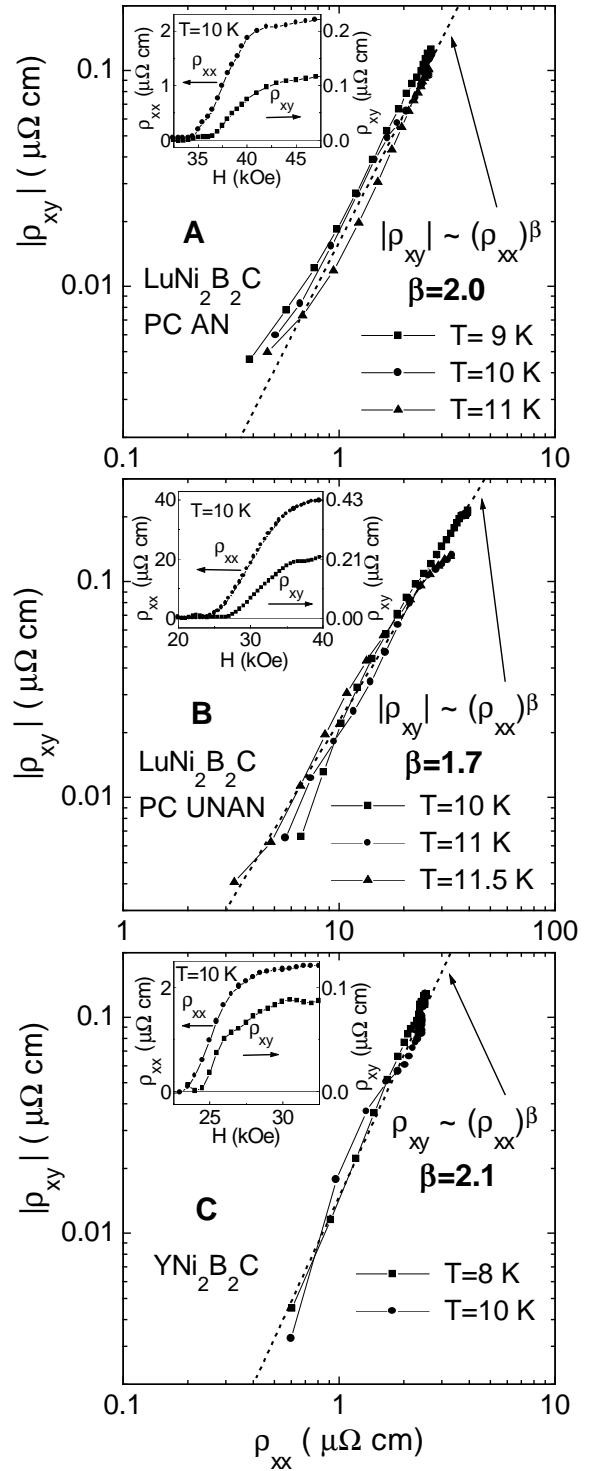}} 
\vspace{0.5pc} 
\caption{$\mid\rho_{xy}\mid$ vs. $\rho_{xx}$ for the annealed (A) and 
unannealed (B) $\rm LuNi_2B_2C$ as well as for the $\rm YNi_2B_2C$ 
(C) samples.  In the insets $\mid\rho_{xy}\mid$ and $\rho_{xx}$ vs.  
magnetic field are simultaneously shown for $T$=10~K.} 
\label{scaling} 
\end{figure}

$\it{In~the~mixed~state}$, the variation of the Hall resistivity with 
magnetic field for both compounds can be described as follows: below 
$T_c$ in low fields there is $\rho_{xy}$=0 as can be seen from the 
$\rho_{xy}(H)$ curves at, e.g., $T$=10~K (Fig.\ 
\ref{rho_xy(H)}A-C ). At higher fields (in the region close to the 
resistive superconducting transition) the Hall resistivity increases 
in the absolute value and gradually reaches the $\rho_{xy}(H)$ curve 
obtained in the normal state at temperatures slightly higher 
than $T_c$.  For $\rm YNi_2B_2C$ the normal state $\rho_{xy}(H)$ 
dependence is very close to linear (see the curve obtained at 
$T$=40~K in Fig.\ \ref{rho_xy(H)}C). At the same time, for both $\rm 
LuNi_2B_2C$ samples the normal state $\rho_{xy}(H)$ dependences have 
a nonlinearity with negative curvature. This nonlinearity, as it was 
pointed out above, is more pronounced for the annealed sample, see 
Fig.\ \ref{rho_xy(H)}A and B.  The Hall resistivity curve 
$\rho_{xy}(H)$ in the mixed state shifts with increasing temperature 
to lower magnetic fields similar to the behavior usually observed 
for the longitudinal resistivity curve $\rho_{xx}(H)$.  
Simultaneously the $\rho_{xy}(H)$ and $\rho_{xx}(H)$ transitions are 
shown in the insets of Fig.\ \ref{scaling} for all samples. Their 
comparison is discussed below.
 
For $\rm LuNi_2B_2C$ as well as $\rm YNi_2B_2C$  no sign reversal of 
$\rho_{xy}(H)$, typical for high-$T_c$ superconductors, was observed 
below $T_c$. The sign of the Hall resistivity is negative in the 
mixed as well as in the normal state.  It should be noted, that for 
high-$T_c$ superconductors not only the hole-like materials, but also 
the electron-like ones usually experience sign reversal of the Hall 
effect (see, e.g., Ref.\ \onlinecite{cagigal}). The behavior of the 
two $\rm LuNi_2B_2C$ samples with significantly different quality 
(the RRR value for the PC UNAN is only $\approx$3) in the mixed state 
is quite similar. Also the behavior of $\rm YNi_2B_2C$ in the mixed 
state is similar to that of $\rm LuNi_2B_2C$. Therefore, the absence 
of the sign reversal of the Hall effect seems to be an intrinsic 
property of the investigated borocarbides.  This result has been 
obtained on polycrystalline samples, but, as has been discussed 
above, the anisotropy of the electronic properties of borocarbides is 
small, and the quality of our annealed samples is high. Therefore, 
this conclusion should remain true also for the borocarbide single 
crystals.  (For high-$T_c$ superconductors, having considerably 
higher anisotropy in electronic properties, sign reversal in the Hall 
effect was observed usually for both poly- and single crystalline 
samples of the same system.)   

In order to understand the absence of sign reversal in the 
$\rho_{xy}$ for the investigated borocarbides, the following physical 
picture of the Hall effect in the mixed state 
\cite{dorsey_pr92,kopnin} can be used:  there are two contributions 
to the Hall conductivity $\sigma_{xy}$ in the superconducting state:

\begin{equation} 
\sigma_{xy}=\sigma_n+\sigma_{sc},
\label{sigma}
\end{equation} 

\noindent
where $\sigma_n$ is connected with normal quasiparticles that 
experience a Lorentz force in the vortex core (it is expected to be 
proportional to $H$) and $\sigma_{sc}$ is an anomalous contribution 
connected with the motion of vortices parallel to the electrical 
current density $\bf j$. In Refs.\ \onlinecite{dorsey_pr92} and 
\onlinecite{kopnin} it was claimed that $\sigma_{sc}\sim1/H$ and 
could have a sign $\em opposite \/$ to that of $\sigma_n$. Therefore, 
at low magnetic fields, the $\sigma_{sc}(H)$ term is more essential 
but at higher fields $\sigma_n(H)$ will be dominant. If
$\sigma_{sc}$ has a different sign than $\sigma_n$ it is possible to 
observe a sign reversal in the Hall effect at $T<T_c$.
\cite{dorsey_pr92,kopnin} Eq.\ (\ref{sigma}) was verified and the 
term $\sigma_{sc}\sim1/H$ was observed, e.g., for YBCO. 
\cite{ginsberg}  For $\rm LuNi_2B_2C$ and $\rm YNi_2B_2C$ the Hall
conductivity decreases with increasing $H$, as can be clearly seen in 
the insets of Fig.\ \ref{rho_xx(H)} for the $\sigma_{xy}$ vs.  $H$ 
curves at $T<T_c$. At the same time it should be pointed out, that 
observed $\sigma_{xy}$ vs. $H$ dependences seem to change more 
rapidly, than $1/H$. A similar behavior was observed for cuprates,
see, e.g., Refs.\ \onlinecite{matsuda_prb95,nagaoka_98}. Therefore, 
the mechanism of the mixed-state Hall effect connected with vortex 
motion seems to work for borocarbides as well.  In the theory 
\cite{kopnin} the sign of the Hall effect in the mixed state is 
determined by the energy derivative $\partial N(0)/\partial \mu$ of 
the density of states $N(0)$ averaged over the Fermi surface.  For a 
complicated Fermi surface which has electron-like and hole-like parts 
the signs of $\sigma_{xy}$ in the normal and in the mixed states may 
be different.  In the phenomenological theory, based on 
Ginsburg-Landau equation and its gauge invariance, \cite{aronov} the 
sign of the Hall conductivity is determined by 
$\partial$ln$T_c/\partial \mu$, where $\mu $ is the chemical 
potential. In any case, the sign of the Hall effect in the mixed 
state depends on the details of the band structure (see also Ref.\ 
\onlinecite{otterlo}).  From our results it follows, that, contrary 
to the case of high-$T_c$ superconductors, the signs of $\sigma_n$ 
and $\sigma_{sc}$ for borocarbides are the same (see the insets of 
Fig.\ \ref{rho_xx(H)}).  This seems to be the reason for the absence 
of sign reversal in the Hall effect in these borocarbides.

In the mixed state two regions concerning the behavior of $\rho_{xy}$ 
and $\rho_{xx}$ can be distinguished. At low magnetic fields both 
$\rho_{xy}$ and $\rho_{xx}$ vanishes.  For higher fields it is 
clearly seen that the scaling behavior $|\rho_{xy}|=A\rho_{xx}^\beta$ 
holds for all three samples (see Fig.\ \ref{scaling}). The values of 
$\beta$ are $2.0\pm$0.1 and $2.1\pm$0.1 for Lu- and Y- based annealed 
samples, respectively. It decreases to 1.7$\pm$0.1 for unannealed 
$\rm LuNi_2B_2C$ sample having a one order of magnitude higher 
resistivity at $T$=17~K. This may be connected with an increase of 
pinning strength for the PC UNAN due to the considerably larger 
concentration of defects leading to the larger resistivity of this 
sample. The decrease of the scaling exponent with increasing 
pinning strength was obtained in the WDT theory \cite{wang_prl94} 
taking into account the backflow current of vortices due to the 
effects of pinning. Another manifestation of the pinning effects, 
predicted by the WDT model, can be seen in the insets of Fig.\ 
\ref{scaling} where the $\rho_{xy}(H)$ and $\rho_{xx}(H)$ curves in 
the superconducting transition region are simultaneously shown. For 
decreasing fields, $\rho_{xx}$ vanishes at definitely lower values of 
$H$ than $|\rho_{xy}|$ for all three samples. The same behavior was 
described by WDT taking into account the effect of pinning. In 
accordance with Ref.\ \onlinecite{wang_prl91}, the Hall resistivity 
can be observed only in the flux flow regime of superconducting 
transition. Prior to the flux flow, the longitudinal resistivity may 
become finite due to flux creep at finite temperature, while the Hall 
resistivity is still zero.  (Vortices in the flux creep regime are 
pinned, and they are only able to creep along the ${\bf j}\times {\bf 
B}$ direction assisted by thermal activation. Creep of flux lines in 
this direction, in accordance with Eq.\ (\ref{josephson}), does not 
lead to a Hall voltage.) From the obtained results it is obvious, 
that pinning effects are considerably important for the mixed state 
Hall effect in the investigated borocarbides.  However, as for 
high-$T_c$ cuprates (see, e.g., Refs.\ 
\onlinecite{jin_98,nagaoka_98}) not only pinning effects will govern 
the mixed-state Hall effect in borocarbides.

\section{Conclusions} 
We have investigated the Hall effect for $\rm LuNi_2B_2C$ and  $\rm 
YNi_2B_2C$ borocarbides in the normal and in the superconducting 
mixed state.  A negative and only slightly temperature-dependent low 
field Hall coefficient was found for both compounds above $T_c$.  The 
value of the Hall coefficient $R_H$ is about one order of magnitude 
smaller, than that resulting from band structure calculations. 
\cite{pickett} A pronounced nonlinearity in the field 
dependence of the Hall resistivity, $\rho_{xy}(H)$, was 
found for $\rm LuNi_2B_2C$ in the normal state below 40~K accompanied 
by a very large positive magnetoresistance MR. An only 
linear $\rho_{xy}(H)$ dependence was observed for $\rm YNi_2B_2C$.  
The possibility of open-orbits formation on the Fermi surface for 
$H\perp c$ is pointed out for $\rm LuNi_2B_2C$ contrary to $\rm 
YNi_2B_2C$.  Measurements of the angular dependence of MR in high 
magnetic fields for Lu- and Y-based single crystals are necessary to 
check the proposed model.  $\em In~the~mixed~state \/$ the behavior 
of Lu- and Y-based borocarbides is quite similar.  Scaling behavior, 
$\rho_{xy}\sim\rho_{xx}^\beta$, was observed but no sign reversal, 
typical for high-$T_c$ superconductors, was found for them.  The 
scaling exponent $\beta$ is $2.0\pm 0.1$ for the annealed $\rm 
LuNi_2B_2C$ sample, it decreases to $1.7\pm 0.1$ for the unannealed 
one, which, in accordance with the WDT theory, can be attributed to 
the increase of the pinning strength.

\acknowledgments
This work was supported by RFBR grant 96-02-00046G and DFG grant 
MU1015/4-1.

\newpage

\widetext
\begin{table}
\caption{Resistivity $\rho_{xx}$ at 300~K and at 17~K, residual 
resistance ratio RRR, critical temperature $T_c$, transition width 
$\Delta T_c$, $|\partial H_{c_2}/\partial T|$ and magnetoresistance 
MR for $\rm LuNi_2B_2C$ and $\rm YNi_2B_2C$ polycrystals (our 
results) and single crystals (data from Refs.\ 
\protect\onlinecite{shulga,rath,fisher_prb97,metlushko})} 
\label{table1} 
\begin{tabular}{ccccccccccc} 
& \multicolumn{3}{c}{Polycrystals} & \multicolumn{7}{c}{Single 
crystals}\\
& Lu & Y & Lu\tablenotemark[1] & Lu & Lu & Lu & Lu & Y & Y 
& Y \\ 
&\multicolumn{2}{c}{annealed} &UN &
[\protect\onlinecite{shulga}] & [\protect\onlinecite{rath}] & 
[\protect\onlinecite{fisher_prb97}] & 
[\protect\onlinecite{metlushko}] & [\protect\onlinecite{shulga}] & 
[\protect\onlinecite{rath}] & [\protect\onlinecite{fisher_prb97}]\\ 
\tableline 
$\rho_{xx}$(300~K) ($\mu\Omega$cm) & 110 & 50 & 134 & 68 & 47 & 
36 & & 108 & 67 & 36\\ 
$\rho_{xx}$(17~K) ($\mu\Omega$cm) & 2.7 & 2.3 & 43 & 2.5 & 1.9 &
1.6 & 1.7 & 2.5 & 3.8 & 2.1\\
RRR & 41 & 22 & 3.1 & 27 & 25 & 23 & & 43 & 18 & 17\\
$T_c$ (K) & 16.7 & 15.5 & 14.7 & 16.5 & 16.1 & 16.0 & 15.8 & 15.7 & 
15.6 & 15.7\\ 
$\Delta T_c$ (K) & 0.27 & 0.44 & 0.6 & 0.2 & 0.25 & & &
0.2 & 0.25 & \\ 
$|\partial H_{c_2}/\partial T|$ (kOe/K)\tablenotemark[2] &
6.8 & 6.0 & 5.8 & 6.4 & 6.2 & & 6.7 & 7.1 & 6.1 & \\ 
MR(20~K,50~kOe) (\% ) & 25 & 10 & 0.7 & & 7.3\tablenotemark[3] & & & & 
7.5\tablenotemark[4] & \\ 
MR(20~K,160~kOe) (\% ) & 90 & 33 & 2.3 & & & & & & & \\ 
\end{tabular} 
\tablenotetext[1]{Unanneled (PC UNAN) $\rm LuNi_2B_2C$ sample.}  
\tablenotetext[2]{For details of the $|\partial H_{c_2}/\partial T|$ 
determination for different samples, see text.}  
\tablenotetext[3]{The value obtained at $T$=20~K and $H$=45~kOe ($H 
\parallel c$).} 
\tablenotetext[4]{The value obtained at $T$=15~K and $H$=45~kOe ($H 
\parallel c$).}  

\end{table}


\begin{references} 

\bibitem[*]{PresAdd}Corresponding author: E-mail:  
narozh@ns.hppi.troitsk.ru

\bibitem[\dagger]{OnLeave}On leave from the Institute for Solid State 
Physics BAS, Sofia, Bulgaria

\bibitem{nagaraj}R. Nagarajan, C. Mazumdar, Z. Hossain, S.K. 
Dhar, K.V. Gopalakrishnan, L.C. Gupta, C. Godart, B.D. Padalia, and 
R. Vijayaraghavan, Phys. Rev. Lett.  {\bf 72}, 274 (1994).

\bibitem{cava}R.J. Cava, H. Takagi, B. Batlog, H.W. Zandbergen, J.J.  
Kraewski, W.F. Peck, Jr., R.B. van Dover, R.J. Felder, T. Siegrist, 
K. Mizuhashi, J.O. Lee, H. Eisaki, S.A. Carter, and S. Uchida, Nature 
{\bf 367}, 252 (1994).

\bibitem{lee}J.I. Lee,T.S. Zao, I.G. Kim, B.I. Min, and S.J. Youn, 
Phys. Rev. {\bf 50}, 4030 (1994).

\bibitem{pickett}W.E. Pickett and D.J. Singh, Phys. Rev. Lett. {\bf 
72}, 3702 (1994).

\bibitem{kim}H. Kim, C.-D. Hwang, and J. Ihm, Phys. Rev. B {\bf 52}, 
4592 (1995).

\bibitem{johnston}E. Johnston-Halperin, J. Fiedler, D.E. Farrell, M.  
Xu, B.K. Cho, P.C. Canfield, D.K. Finnemore, and D.C. Johnston, Phys. 
Rev. B {\bf 51}, 12852 (1995).

\bibitem{fisher_prb97}I.R. Fisher, J.R. Cooper, and P.C. Canfield, 
Phys. Rev. B {\bf 56}, 10820 (1997).

\bibitem{metlushko}V. Metlushko, U. Welp, A. Koshelev, I. Aranson, 
G.W. Crabtree, and P.C. Canfield, Phys. Rev.  Lett.  {\bf 79}, 1738 
(1997).

\bibitem{rath}K.D.D. Rathnayaka, A.K. Bhatnagar, D.G.  Naugle, P.C. 
Canfield, and B.K. Cho, Phys.  Rev.  B {\bf 55}, 8506 (1997).

\bibitem{eisaki}H. Eisaki, H. Takagi, R.J. Cava, B. Batlogg, J.J.  
Krajewski, W.F. Peck, Jr, K.Mazuhashi, J.O. Lee, and S. Uchida, Phys.  
Rev. B {\bf 50}, 647 (1994).

\bibitem{dewilde}Y. de Wilde,M. Iavaron, U. Welp, V. Metlushko, A.E.  
Koshelev, I. Aranson, G.W. Crabtree, and P.C. Canfield, Phys.  Rev.  
Lett.  {\bf 78}, 4273 (1997).

\bibitem{eskildsen}M. R. Eskildsen, P.L. Gammel, B.P. Barder, U.  
Yaron, A.P. Ramirez, D.A. Huse, D.J. Bishop, C. Bolle, C.M. Lieber, 
S.  Oxx, S. Sridhar, N.H. Andersen, K. Mortensen, and P.C. Canfield, 
Phys.  Rev.  Lett. {\bf 78}, 1968 (1997). 

\bibitem{yethiraj_prl97}M. Yethiraj, D.McK. Paul, C.V. Tomy, and E.M. 
Forgan, Phys.  Rev.  Lett. {\bf 78}, 4849 (1997).

\bibitem{eskildsen_prlj97}M. R. Eskildsen, P.L. Gammel, B.P. Barder, 
A.P. Ramirez, D.J. Bishop, N.H. Andersen, K. Mortensen, C.A. Bolle, 
C.M.  Lieber, and P.C.  Canfield, Phys.  Rev.  Lett. {\bf 79}, 487 
(1997).
 
\bibitem{fischer}I.R. Fisher, J.R. Cooper, and R.J. Cava, Phys. Rev.
B {\bf 52}, 15086 (1995).

\bibitem{narozh_jltp}V.N. Narozhnyi, V.N. Kochetkov, A.V.  
Tsvyashchenko, and L.N. Fomicheva, J. Low Temp. Phys. {\bf 105}, 1647 
(1996).

\bibitem{mandal}P. Mandal and K. Winzer, Solid State Communs. {\bf 
103}, 679 (1997).

\bibitem{narozh_jac}V.N. Narozhnyi, V.N. Kochetkov, A.V.  
Tsvyashchenko, and L.N. Fomicheva, J. Alloys and Compouds {\bf 
275-277}, 484 (1998).

\bibitem{josephson}B.D. Josephson, Phys. Lett. {\bf 16}, 242 (1965).

\bibitem{galffy}M. Galffy and E. Zirngiebl, Solid State Communs.  
{\bf 68}, 929 (1988).

\bibitem{ri}H.-C. Ri, R. Gross, F. Gollnik, A. Beck, R.P.  Huebener, 
P. Wagner and H. Adrian, Phys.  Rev.  B {\bf 50}, 3312 (1994).

\bibitem{samoilov_93}A.V. Samoilov, Phys. Rev. Lett. {\bf 71}, 617 
(1993).

\bibitem{budhani}R.C. Budhani, S.H. Liou and Z.X. Cai, Phys. Rev.  
Lett. {\bf 71}, 621 (1993).

\bibitem{hagen_prb93}S.J. Hagen, A.W. Smith, M. Rajeswari, J.L.  
Peng, Z.Y. Li, R.L. Green, S.N. Mao, X.X. Xi, S. Bhattacharya, Q. Li, 
and C.J. Lobb, Phys. Rev. B {\bf 47}, 1064 (1993).

\bibitem{cagigal}M. Cagigal, J. Fontcuberta, M.A. Crusellas, J.L.  
Vicent and S. Pinol, Physica C {\bf 248}, 155 (1995).

\bibitem{li}K. Li, Y. Zhang, and H. Adrian, Phys. Rev. B {\bf 53}, 
8608 (1996).

\bibitem{wang_prl97}H.C. Yang, L.M. Wang, and H.E. Horng, Phys. Rev. B 
{\bf 56}, 99 (1997).

\bibitem{yang_pr97}L.M. Wang, H.C. Yang, and H.E. Horng, Phys. Rev. 
Lett. {\bf 78}, 527 (1997).

\bibitem{parks}Y.B. Kim and M.J. Stephen, in {\it Superconductivity}, 
ed. by R.D. Parks (Marcel Dekker, New York, 1969).

\bibitem{smith}A.W. Smith, T.W. Clinton, C.C. Tsuei, and C.J. Lobb, 
Phys. Rev. B {\bf 49}, 12927 (1994).

\bibitem{bhat}S. Bhattacharya, M.J. Higgins, and T.V. Ramakrishan, 
Phys.  Rev. Lett. {\bf 73}, 1699 (1994).

\bibitem{BS}J. Bardeen and M.J. Stephen, Phys. Rev. {\bf 140}, A1197 
(1965).

\bibitem{NV}P. Nozi\`eres and W.F. Vinen, Philos. Mag. {\bf 14}, 667 
(1966).
 
\bibitem{wang_prl91}Z.D. Wang and C.S. Ting, Phys. Rev. Lett.  {\bf 
67}, 3618 (1991).

\bibitem{dorsey_pr92}A.T. Dorsey, Phys. Rev. B {\bf 46}, 8376 (1992).

\bibitem{kopnin}N.B. Kopnin, B.I. Ivlev, and V.A. Kalatsky, J. Low 
Temp.  Phys. {\bf 90}, 1 (1993).

\bibitem{aronov}A.G. Aronov, S. Hikami and A.I. Larkin, Phys. Rev. B 
{\bf 51}, 3880 (1995).
 
\bibitem{otterlo}A. van Otterlo, M. Feigel'man, V. Geshkenbein and 
G. Blatter Phys. Rev. Lett. {\bf 75}, 3736 (1995).

\bibitem{khomskii}D.I. Khomskii and A. Freimuth, Phys. Rev. Lett {\bf 
75}, 1384 (1995).
  
\bibitem{jin_98}R. Jin and H.R. Ott, Phys. Rev. B {\bf 57}, 13872 
(1998).
 
\bibitem{nagaoka_98}T. Nagaoka, Y. Matsuda, H. Obara, A. Sawa, T. 
Terashima, I. Chong, M. Takano and M. Suzuki, Phys.  Rev.  Lett.  
{\bf 80}, 3594 (1998).

\bibitem{luo}J. Luo, T.P. Orlando, J.M. Graybeal, X.D. Wu, and R.  
Muenchausen, Phys. Rev. Lett. {\bf 68}, 690 (1992).

\bibitem{rice}J.P. Rice, N. Rigakis, D.M. Ginsberg, and J.M. Mochel, 
Phys. Rev. B {\bf 46}, 11050 (1992). 

\bibitem{samoilov_pr94}A.V. Samoilov, Z.G. Ivanov, and L.G.  
Johansson, Phys. Rev. B {\bf 49}, 3667 (1994).

\bibitem{okuma}S. Okuma and N. Kokubo, Phys. Rev. B {\bf 56}, 410 
(1997).

\bibitem{dorsey_fischer}A.T. Dorsey and M.P.A. Fisher, Phys. Rev.  
Lett.  {\bf 68}, 694 (1992).

\bibitem{vinokur_prl93}V.M. Vinokur, V.B. Geshkenbein, M.V.  
Feigel'man, and G. Blatter, Phys.  Rev.  Lett.  {\bf 71}, 1242 (1993).

\bibitem{wang_prl94}Z.D. Wang, J. Dong, and C.S. Ting, Phys. Rev.  
Lett.  {\bf 72}, 3875 (1994).

\bibitem{kang}W.N. Kang, D.H. Kim, S.Y. Shim, J.H. Park, T.S. Hahn, 
S.S. Choi, W.C. Lee, J.D. Hettinger, K.E. Gray, and B. Glagola, Phys. 
Rev. Lett. {\bf 76}, 2993 (1996). 

\bibitem{samoilov_prl95}A.V. Samoilov, A. Legris, F.  
Rullier-Albenque, P. Lejay, S. Bouffard, Z.G. Ivanov and L.-G.  
Johansson, Phys. Rev. Lett.  {\bf 74}, 2351 (1995).

\bibitem{morgoon}V.N. Morgoon, V.A. Shklovskij, V. Bindilatti, A.V. 
Bondarenko, R.F. Jardim, C.C. Becerra, C.Y. Shigue, A.V. Sivakov, J.  
Low Temp.  Phys.  {\bf 105}, 963 (1996).

\bibitem{maki_prb98}G. Wang and K. Maki, Phys. Rev. B {\bf
58}, 6493 (1998).

\bibitem{narozh_ssc}V.N. Narozhnyi, J. Freudenberger, V.N. 
Kochetkov, K.A. Nenkov, G. Fuchs, and K.-H. M\"uller,  
Solid State Commun. {\bf 109}, 549 (1999).
 
\bibitem{muller_96}K. Eversmann, A. Handstein, G. Fuchs, L.  Gao, and 
K.-H. M\"uller, Physica C {\bf 266}, 27 (1996).

\bibitem{shulga}S.V. Shulga, S.-L. Drechsler, G. Fuchs, K.-H.  
M\"uller, K. Winzer, M. Heinecke, and K. Krug, Phys. Rev. Lett.  {\bf 
80}, 1730 (1998).

\bibitem{aleks}N.E. Alekseevskii, V.I. Nizhankovskii, V.N. Narozhnyi, 
E.P. Khlybov, and A.V. Mitin, J. Low Temp. Phys. {\bf 64}, 87 (1986).

\bibitem{gollnik_prb98}F. Gollnik and M. Naito, Phys. Rev. B {\bf
58}, 11734 (1998). 

\bibitem{nguyen}L.H. Nguyen, G. Goll, E. Steep, A.G.M. Jansen, P.  
Wyder, O. Jepsen, M. Heinecke, and K. Winzer, J. Low Temp. Phys.  
{\bf 105}, 1653 (1996).

\bibitem{lifshitz}I.M. Lifshitz, M.Ya. Azbel', and M.I. Kaganov, in 
{\it Electron Theory of Metals}, (Consultants Bureau, NY, 1973).
 
\bibitem{ginsberg}D.M. Ginsberg and J.T. Manson, Phys. Rev. B 
{\bf 51}, 515 (1995).

\bibitem{matsuda_prb95}Y. Matsuda, T. Nagaoka, G. Suzuki, K. 
Kumagai, M. Suzuki, M. Machida, M. Sera, M. Hiroi, and N. Kobayashi, 
Phys.  Rev. B {\bf 52}, 15749 (1995).

\end{references}
\end{document}